\documentclass[aps,prl,twocolumn,groupedaddress,longbibliography]{revtex4-1}

\usepackage{graphicx}

\begin{document}
	

\title{Coupling Rydberg atoms to microwave fields in a superconducting coplanar waveguide resonator}


\author{A. A. Morgan}
\author{S. D. Hogan}
\affiliation{Department of Physics and Astronomy, University College London, Gower Street, London WC1E 6BT, United Kingdom}
	

\date{\today}
	
\begin{abstract}
Rydberg helium atoms traveling in pulsed supersonic beams have been coupled to microwave fields in a superconducting coplanar waveguide (CPW) resonator. The atoms were initially prepared in the 1s55s\,$^3$S$_1$ Rydberg level by two-color two-photon laser excitation from the metastable 1s2s\,$^3$S$_1$ level. Two-photon microwave transitions between the 1s55s\,$^3$S$_1$ and 1s56s\,$^3$S$_1$ levels were then driven by the 19.556~GHz third-harmonic microwave field in a quarter-wave CPW resonator. This superconducting microwave resonator was fabricated from niobium nitride on a silicon substrate and operated at temperatures between 3.65 and 4.30~K. The populations of the Rydberg levels in the experiments were determined by state-selective pulsed electric field ionization. The coherence of the atom-resonator coupling was studied by time-domain measurements of Rabi oscillations. 
\end{abstract}
	
	
\maketitle
	
Hybrid quantum systems, composed of individual components with complementary characteristics, e.g., coherence times, state manipulation rates, available transition frequencies, or scalability, are of interest in a range of areas, from quantum computation and communication to quantum sensing (see, e.g., Refs.~\cite{wallquist09a,xiang13a,kurizki15a}). Initial proposals to use gas-phase atoms or molecules as quantum memories and optical interfaces for solid-state quantum processors~\cite{rabl06a} were conceived to combine (1) the long-coherence times of transitions between rotational states in polar molecules with electric dipole moments of $\sim1~e\,a_0$ ($e$ and $a_0$ are the electron charge and Bohr radius, respectively), and (2) the scalability and fast quantum-state manipulation achievable in solid-state superconducting circuits~\cite{wallraff04a}. Coupling between these systems was envisaged using chip-based superconducting coplanar waveguide (CPW) microwave resonators~\cite{wallraff04a,goppl08a}. Atoms in Rydberg states with high principal quantum number, $n$, are also well suited to this approach to hybrid microwave cavity quantum electrodynamics (QED), and have played a central role in many seminal cavity QED experiments with three-dimensional microwave cavities (see, e.g., Ref.~\cite{haroche13a}, and references therein). In the hybrid cavity QED setting, the large electric dipole moments associated with microwave transitions between Rydberg states (typically $>1000~e\,a_0$ for $n\gtrsim30$~\cite{gallagher94a}) offer higher single-particle atom-resonator coupling strengths than ground-state molecules. These can be achieved while maintaining a wide range of available transition frequencies for applications in optical-to-microwave photon conversion~\cite{kiffner16a,han18a}, and -- in the case of circular Rydberg states -- lifetimes $>10$~ms for quantum memories.

The development of hybrid approaches to cavity QED with Rydberg atoms and microwave circuits began with experiments in which $|n\mathrm{s}\rangle\rightarrow|n\mathrm{p}\rangle$ transitions were driven by microwave fields in copper CPWs~\cite{hogan12a}. This highlighted effects of stray electric fields emanating from the surfaces of the microwave circuits that must be considered in these experiments performed in cryogenic environments. These studies were performed using helium (He) atoms to minimize adsorption, and associated time-dependent changes in stray electric fields emanating from the surfaces~\cite{hattermann12a}. Subsequently, proposals were put forward to implement quantum gates~\cite{pritchard14a,sarkany15a}, quantum-state transfer~\cite{patton13a,sarkany18a}, and microwave-to-optical photon conversion~\cite{gard17a,petrosyan19a}, and to access new regimes of light-matter interactions~\cite{calajo17a} in hybrid systems composed of Rydberg atoms and superconducting circuits. Transitions between Rydberg states have have been driven by microwave fields propagating in superconducting coplanar waveguides~\cite{hermann14a,thiele15a}, and ground-state rubidium atoms have been coupled to microwave fields in superconducting CPW resonators~\cite{hattermann17a}. However, up to now coupling between Rydberg atoms and microwave fields in CPW resonators has not been reported. 

Here we present the results of experiments in which Rydberg atoms, in particular He atoms initially prepared in the 1s55s\,$^3$S$_1$ ($|55\mathrm{s}\rangle$) level, have been coupled to microwave fields in a chip-based superconducting CPW resonator, with a mode volume $\sim10^{-6}$ times smaller than that of any resonator used previously in microwave cavity QED experiments with atoms in Rydberg states~\cite{gleyzes07a,garcia19a}. These fields, at frequencies close to 19.556~GHz, were resonant with the 1s55s\,$^3$S$_1\rightarrow$1s56s\,$^3$S$_1$ ($|55\mathrm{s}\rangle\rightarrow|56\mathrm{s}\rangle$) two-photon transition. These $|n\mathrm{s}\rangle$ Rydberg states are appropriate for directly interfacing microwave photons in the superconducting circuit with optical photons for long-distance information transfer. The methods employed could be implemented with long-lived circular Rydberg states in He~\cite{morgan18a} for use as quantum memories. This hybrid cavity QED system is of interest in accessing of new regimes of light-matter interactions~\cite{calajo17a}, and for nondemolition detection~\cite{gleyzes07a,garcia19a} of cold trapped Rydberg molecules~\cite{hogan09a}, and Rydberg positronium atoms~\cite{wall15a}.  

\begin{figure}
\begin{center}
\includegraphics[width = 0.425\textwidth, angle = 0, clip=]{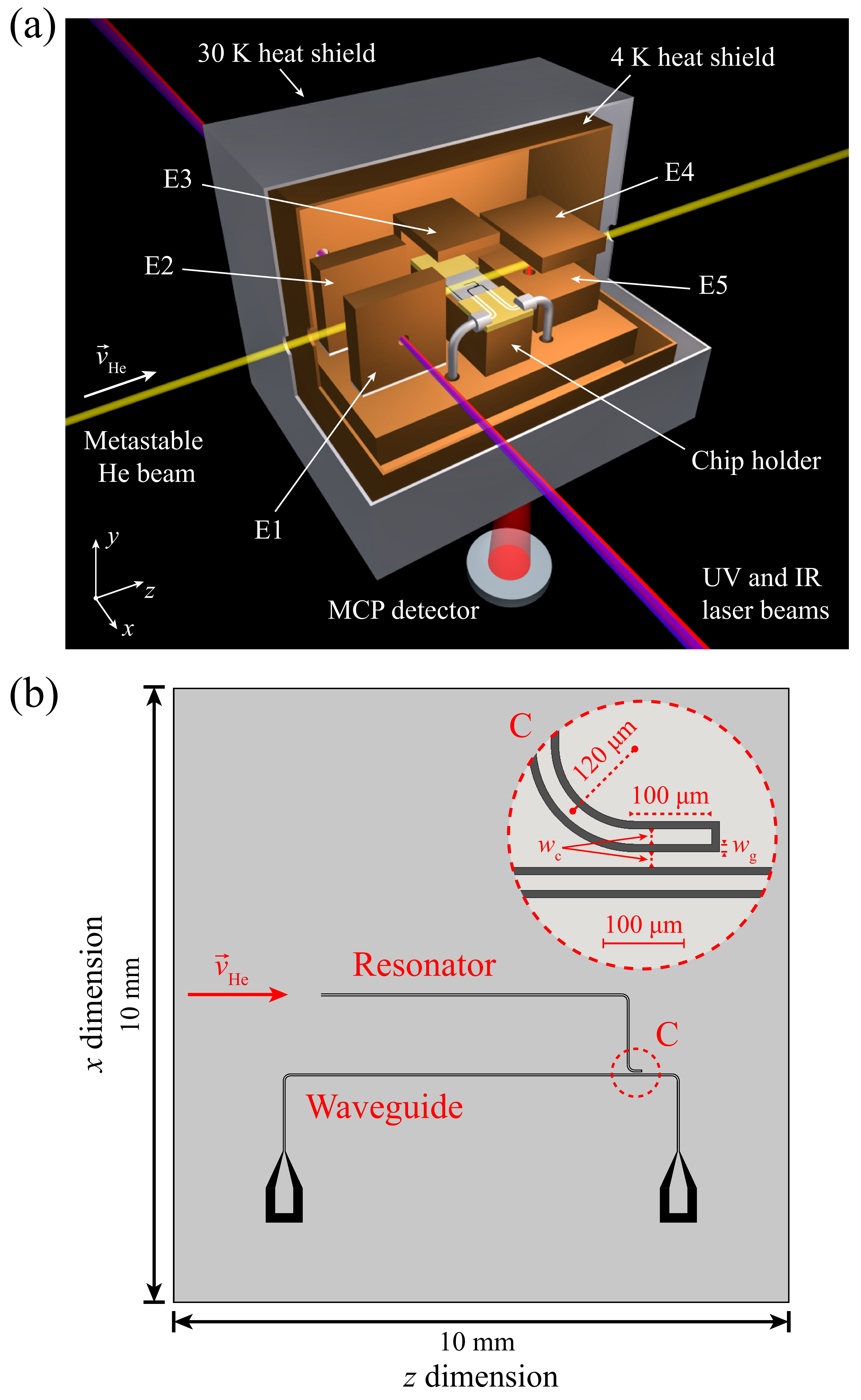}
\caption{(a) Schematic diagram of the experimental apparatus (not to scale). For clarity parts of the heat shields and electrode E3 have been omitted. (b) Diagram of the $10\times10$~mm NbN chip containing a U-shaped CPW and an L-shaped $\lambda/4$ CPW resonator along which the atomic beam propagated. The inset in (b) represents an expanded view of the coupling region, C, between the waveguide and resonator.}
\label{fig1}
\end{center}
\end{figure}

A schematic diagram of the experimental apparatus is displayed in Fig.~\ref{fig1}(a). A pulsed supersonic beam of He atoms ($\langle v_{\mathrm{He}}\rangle = 2000$~m/s) in the metastable 1s2s\,$^3$S$_1$ level was generated in an electric discharge at the exit of a pulsed valve (repetition rate 50~Hz)~\cite{halfmann00a}. After passing through a 2-mm-diameter skimmer and a charged-particle filter, the atoms entered through apertures of 4 and 3~mm diameter in two heat shields, maintained at 30~K and $<4$~K, respectively, and into the cryogenic region. Between two electrically isolated parallel copper blocks, E1 and E2 in Fig.~\ref{fig1}(a), the atoms were excited to Rydberg states using the $1\mathrm{s}2\mathrm{s}\,^3\mathrm{S}_1\rightarrow1\mathrm{s}3\mathrm{p}\,^3\mathrm{P}_2\rightarrow1\mathrm{s}55\mathrm{s}\,^3\mathrm{S}_1$ two-color two-photon excitation scheme~\cite{hogan18a}. This was driven with focused ($\sim100~\mu$m FWHM beam waist) co-propagating continuous wave (CW) laser radiation at wavelengths of 388.975~nm and 786.817~nm, respectively, in a pulsed electric field of 0.3~V/cm. This field was switched on for 1.25~$\mu$s generating 2.5-mm-long bunches of Rydberg atoms. 

After excitation the atoms traveled 25~mm to the grounded end of a $\lambda/4$ niobium nitride (NbN) (critical temperature $T_{\mathrm{c}}=12.1$~K) superconducting CPW microwave resonator [Fig.~\ref{fig1}(b)]. This resonator was fabricated by photolithography of a 100-nm-thick NbN film on a 10~mm$\times$10~mm silicon chip. The resonator was L-shaped with a length of 6.335~mm, and center-conductor and insulating gap widths of $w_{\mathrm{c}}=20~\mu$m and $w_{\mathrm{g}}=10~\mu$m, respectively  [see inset in Fig.~\ref{fig1}(b)]. At the open end, the resonator was capacitively coupled to a U-shaped superconducting CPW. The center conductor and ground planes of this CPW were electrically connected to a larger copper CPW on a printed circuit board [Fig.~\ref{fig1}(a)]. The resonator was operated at temperatures close to $T_{\mathrm{CPW}}=3.65$~K for which its third-harmonic frequency was $\nu_3=19.559\,41$~GHz, loaded quality factor was $Q_{\mathrm{loaded}}=2280$, and build-up/ring-down time was $\tau=18.4$~ns. A $\lambda/4$ CPW resonator geometry, with one grounded end and one open end, was chosen to minimize charge buildup on the center conductor. 

As the Rydberg atoms traveled along the 4.858-mm-long section of the resonator aligned with the axis of propagation of the atomic beam [see Fig.~\ref{fig1}(b)], the resonant enhancement in the local microwave field was exploited to drive two-photon $|55\mathrm{s}\rangle\rightarrow|56\mathrm{s}\rangle$ transitions at $\nu_{55\mathrm{s},56\mathrm{s}}/2=19.556\,499$~GHz. 16~mm beyond the NbN chip the atoms were detected by state-selective pulsed electric field ionization by applying a slowly-rising pulsed ionization potential of $-125$~V to E4 while E5 was maintained at 0~V. Ionized electrons were accelerated through a 5-mm-diameter aperture in E5 and out of the cryogenic region to a room-temperature microchannel plate (MCP) detector [Fig.~\ref{fig1}(a)].

\begin{figure}
\begin{center}
\includegraphics[width = 0.45\textwidth, angle = 0, clip=]{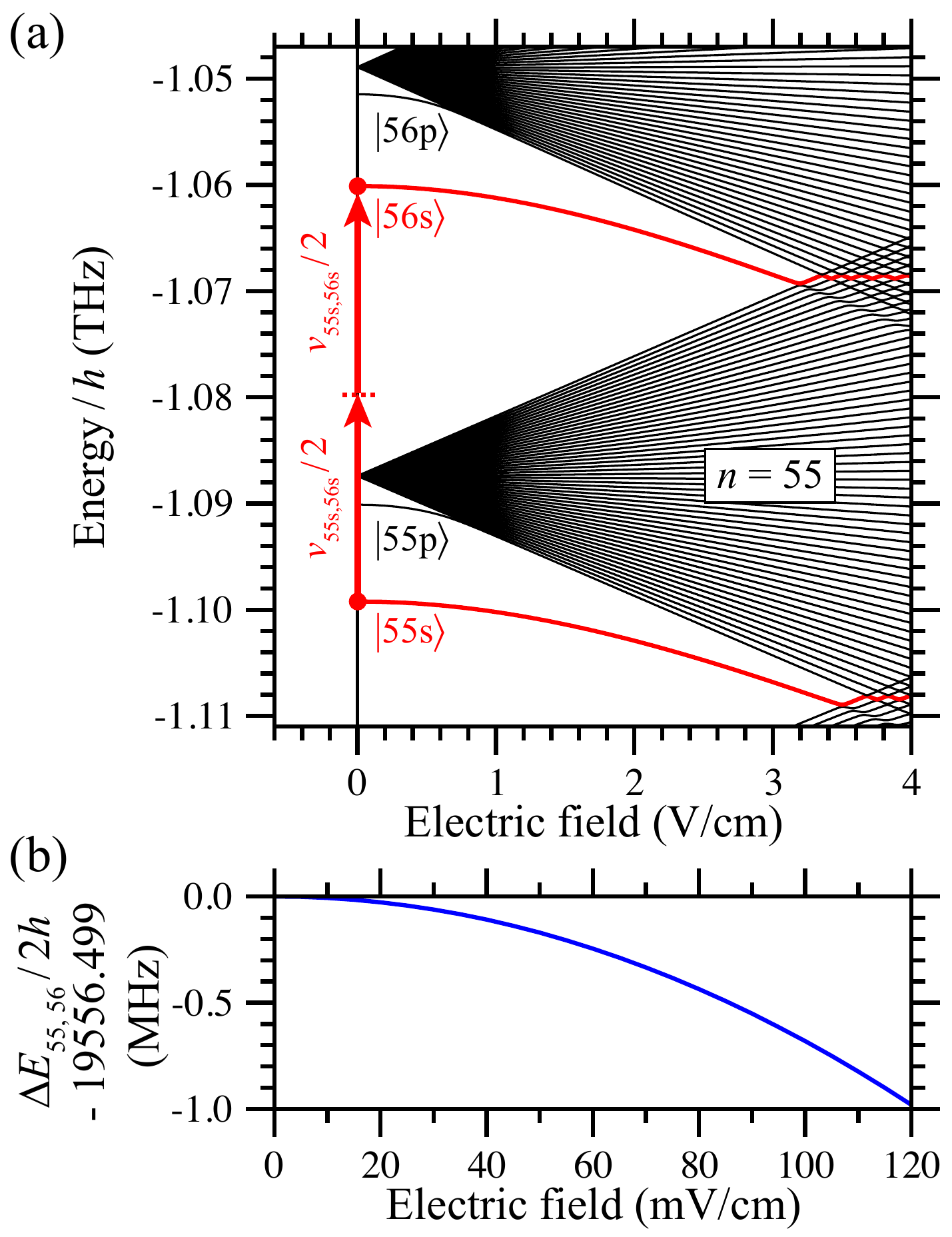}
\caption{(a) Stark map of the triplet Rydberg states in He with $m_{\ell}=0$. The two-photon $|55\mathrm{s}\rangle\rightarrow|56\mathrm{s}\rangle$ transition is indicated by the vertical red arrows. (b) Electric-field dependence of the $|55\mathrm{s}\rangle\rightarrow|56\mathrm{s}\rangle$ two-photon transition frequency.}
\label{fig2}
\end{center}
\end{figure}

The triplet $|55\mathrm{s}\rangle$ ($|56\mathrm{s}\rangle$) Rydberg state in He has a quantum defect of 0.296\,669\,3 (0.296\,668\,8)~\cite{drake99a}. As seen in Fig.~\ref{fig2}(a) this state exhibits a quadratic Stark shift in weak electric fields, with a static electric dipole polarizability of 1.955\,45~GHz/(V/cm)$^2$ [2.202\,11~GHz/(V/cm)$^2$]. The small difference between the $|55\mathrm{s}\rangle$ and $|56\mathrm{s}\rangle$ polarizabilities means that the $|55\mathrm{s}\rangle\rightarrow|56\mathrm{s}\rangle$ transition has a low sensitivity to residual uncanceled stray electric fields, and electric field inhomogeneities. This can be seen from the Stark shift of $\nu_{55\mathrm{s},56\mathrm{s}}/2$ in fields up to 120~mV/cm in Fig.~\ref{fig2}(b).

\begin{figure}
\begin{center}
\includegraphics[width = 0.4\textwidth, angle = 0, clip=]{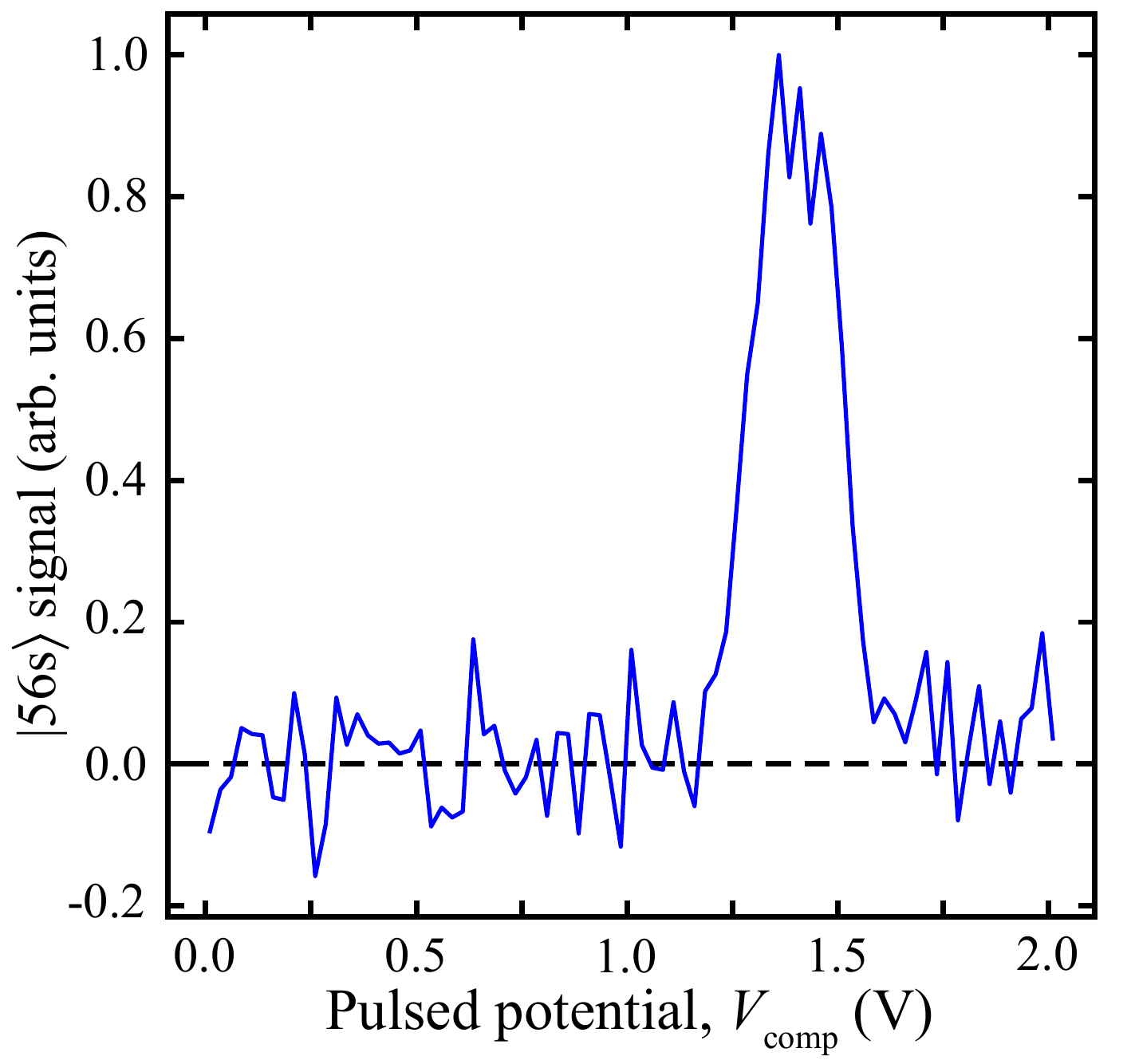}
\caption{Dependence of the population transfer to the $|56\mathrm{s}\rangle$ state on the potential, $V_{\mathrm{comp}}$, applied to electrode E3 to compensate stray electric fields close to the NbN chip surface. In recording this data $T_{\mathrm{CPW}}=3.65$~K, and the resonator was driven with a microwave field at the frequency $\nu_{55\mathrm{s},56\mathrm{s}}/2$.}
\label{fig3}
\end{center}
\end{figure}

To couple Rydberg atoms to the microwave field in the CPW resonator the apparatus in Fig.~\ref{fig1}(a) was cooled to cryogenic temperatures following a procedure similar to that in Ref.~\cite{thiele14a}. Throughout the cooling process the NbN chip was maintained at a higher temperature than the other components to minimize adsorption. The resonator was then stabilized to $T_{\mathrm{CPW}}=3.65$~K so that $\nu_3=19.559\,41$~GHz. When the Rydberg atoms passed over the first antinode of the microwave field, $13.25~\mu$s after photoexcitation, a 500-ns-long pulse of microwave radiation at the frequency $\nu_{55\mathrm{s},56\mathrm{s}}/2$, and a CW output power at the source of $P_{\mathrm{source}}=2$~mW (the attenuation between the source and the CPW on the NbN chip was $\sim-22.5$~dB), was injected into the resonator to drive the $|55\mathrm{s}\rangle\rightarrow|56\mathrm{s}\rangle$ transition. To compensate stray electric fields above the NbN chip, the population transfer from the $|55\mathrm{s}\rangle$ to the $|56\mathrm{s}\rangle$ state, driven by this resonator field, was monitored while a pulsed potential, $V_{\mathrm{comp}}$, applied to E3 was adjusted. The resulting data in Fig.~\ref{fig3} indicate that maximal population transfer occurred for $V_{\mathrm{comp}}\simeq+1.45$~V. With this potential applied to E3, the electric field at the position of the atoms was therefore minimized. This compensation potential was applied for all subsequent measurements. 

\begin{figure}
\begin{center}
\includegraphics[width = 0.4\textwidth, angle = 0, clip=]{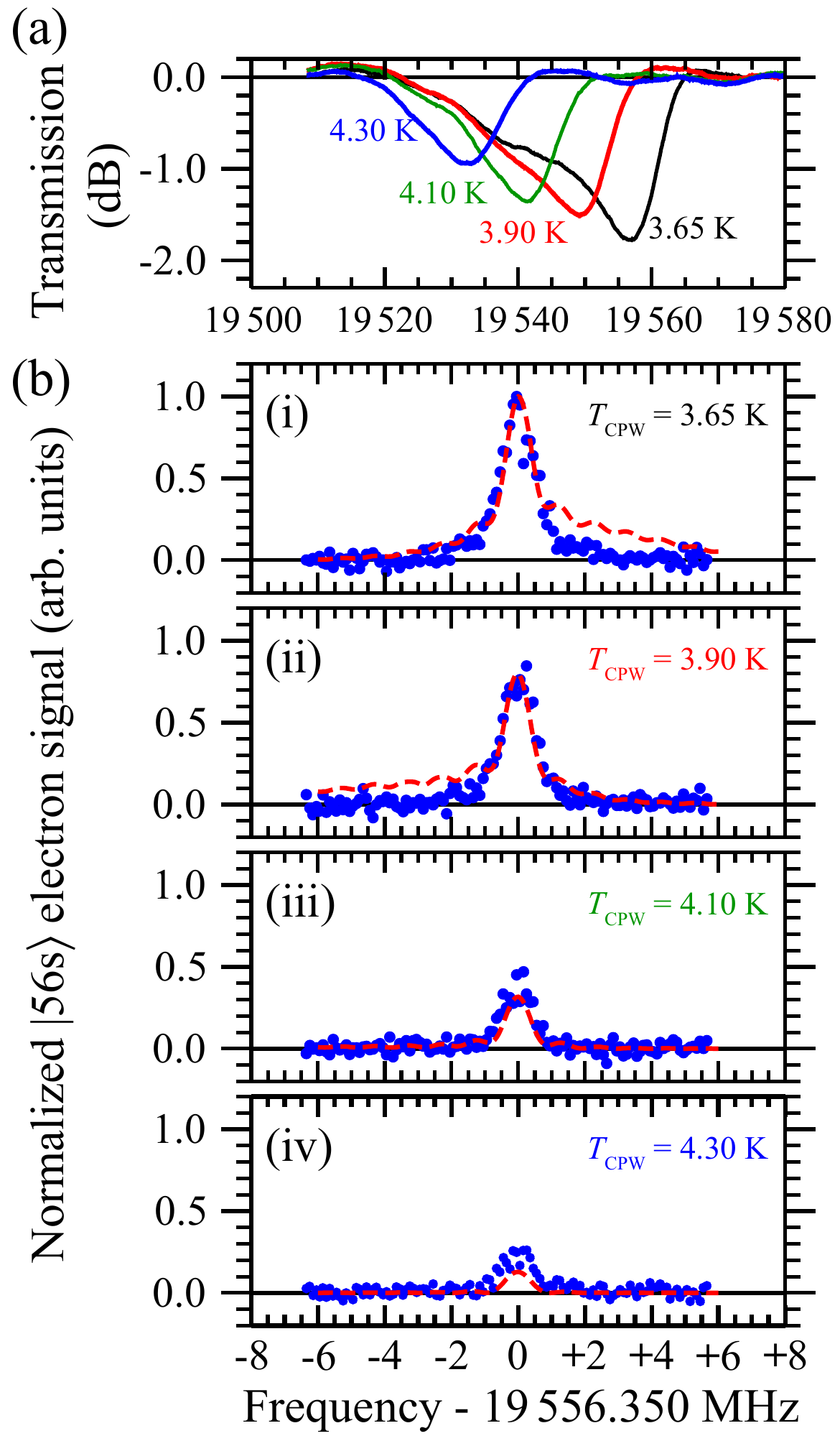}
\caption{(a) Temperature dependence of resonator line shape recorded by monitoring the CPW transmission. (b-i)--(b-iv) Spectra of the $|55\mathrm{s}\rangle\rightarrow|56\mathrm{s}\rangle$ two-photon transition recorded for $P_{\mathrm{source}}=2$~mW and the values of $T_{\mathrm{CPW}}$ indicated. Experimentally recorded (calculated) data are indicated by the blue points (dashed red curves).}
\label{fig4}
\end{center}
\end{figure}

The coupling of the atoms to the microwave field in the resonator was confirmed by recording spectra for a range of detunings of $\nu_3$ from $\nu_{55\mathrm{s},56\mathrm{s}}/2$. The value of $\nu_3$ is strongly dependent on $T_{\mathrm{CPW}}$. This can be seen from the spectra of microwave transmission through the U-shaped CPW on the NbN chip in Fig.~\ref{fig4}(a). The measured resonances exhibit asymmetric line shapes as a result of interference between the transmitted microwave field, and that scattered by the resonator. The values of $\nu_3$, and the internal, $Q_{\mathrm{int}}$, and loaded, $Q_{\mathrm{loaded}}$, quality factors of the resonator, i.e., the quality factor of the isolated resonator, and the modified quality factor arising as a result of the coupling to the waveguide, determined from these spectra using the circle-fit method~\cite{probst15a} are listed in Table~\ref{tab1}.
\begin{table}
\begin{tabular*}{8cm}{@{\extracolsep{\fill}} c c c c}
\hline
\hspace*{0.4cm}$T_{\mathrm{CPW}}$ (K) & $\nu_3$ (GHz) & $Q_{\mathrm{int}}$ & $Q_{\mathrm{loaded}}$\hspace*{0.4cm}\\
\hline
\hspace*{0.4cm}3.65 & 19.559\,41 & $2900\pm135$ & $2280\pm90$\hspace*{0.4cm}\\
\hspace*{0.4cm}3.90 & 19.551\,11 & $2635\pm130$ & $2140\pm90$\hspace*{0.4cm}\\
\hspace*{0.4cm}4.10 & 19.542\,61 & $2650\pm110$ & $2200\pm80$\hspace*{0.4cm}\\
\hspace*{0.4cm}4.30 & 19.533\,99 & $2720\pm115$ & $2390\pm90$\hspace*{0.4cm}\\
\hline
\end{tabular*}
\caption{Temperature dependence of $\nu_3$, and the internal, $Q_{\mathrm{int}}$, and loaded, $Q_{\mathrm{loaded}}$, quality factors of the CPW resonator determined from the spectral function representing the microwave transmission through the waveguide displayed in Fig.~\ref{fig4}(a).}\label{tab1}
\end{table}
For $T_{\mathrm{CPW}}=3.65$~K, $\nu_3=19.559\,41$~GHz and the resonator resonance had a FWHM spectral width of $8.5$~MHz. A microwave spectrum of the $|55\mathrm{s}\rangle\rightarrow|56\mathrm{s}\rangle$ transition recorded under these conditions, with 500-ns-long microwave pulses injected into the resonator and $P_{\mathrm{source}}=2$~mW, is displayed in Fig.~\ref{fig4}(b-i) (blue points). The centroid of the observed spectral line lies at 19.556\,350~GHz, and its width is $\Delta\nu_{\mathrm{FWHM}}\simeq1.4$~MHz. Comparing this measured transition frequency with $\nu_{55\mathrm{s},56\mathrm{s}}/2$, and the data in Fig.~\ref{fig2}(b), indicates that the average stray electric field at the position of the atoms above the resonator was $<50$~mV/cm. When $\nu_3$ was detuned from $\nu_{55\mathrm{s},56\mathrm{s}}/2$ the enhancement in the microwave field on resonance with the atomic transition at the position of the atoms was reduced. This resulted in a corresponding reduction in the intensity of the spectral features for $T_{\mathrm{CPW}}=3.9$, 4.1 and 4.3~K in Figs.~\ref{fig4}(b-ii) to (b-iv). The reduction in $|56\mathrm{s}\rangle$ signal as $T_{\mathrm{CPW}}$ was increased confirms that the atoms were coupled to the microwave field enhanced within the CPW resonator. The temperature dependence of the intensity of the atomic transition is in good agreement with that calculated by considering the coherent evolution of an ensemble of atoms with 100-$\mu$m-FWHM Gaussian spatial distributions in the $x$ and $y$ dimensions, centered 100~$\mu$m above the resonator center conductor [dashed red curves in Fig.~\ref{fig4}(b)]. In these calculations the power spectrum within the resonator was described by a Lorentzian function with the values of $\nu_3$ and $Q_{\mathrm{loaded}}$ in Table~\ref{tab1}. The spatial distribution of the microwave field above the resonator was determined by finite element methods~\cite{morgan18a}. 

\begin{figure}
\begin{center}
\includegraphics[width = 0.4\textwidth, angle = 0, clip=]{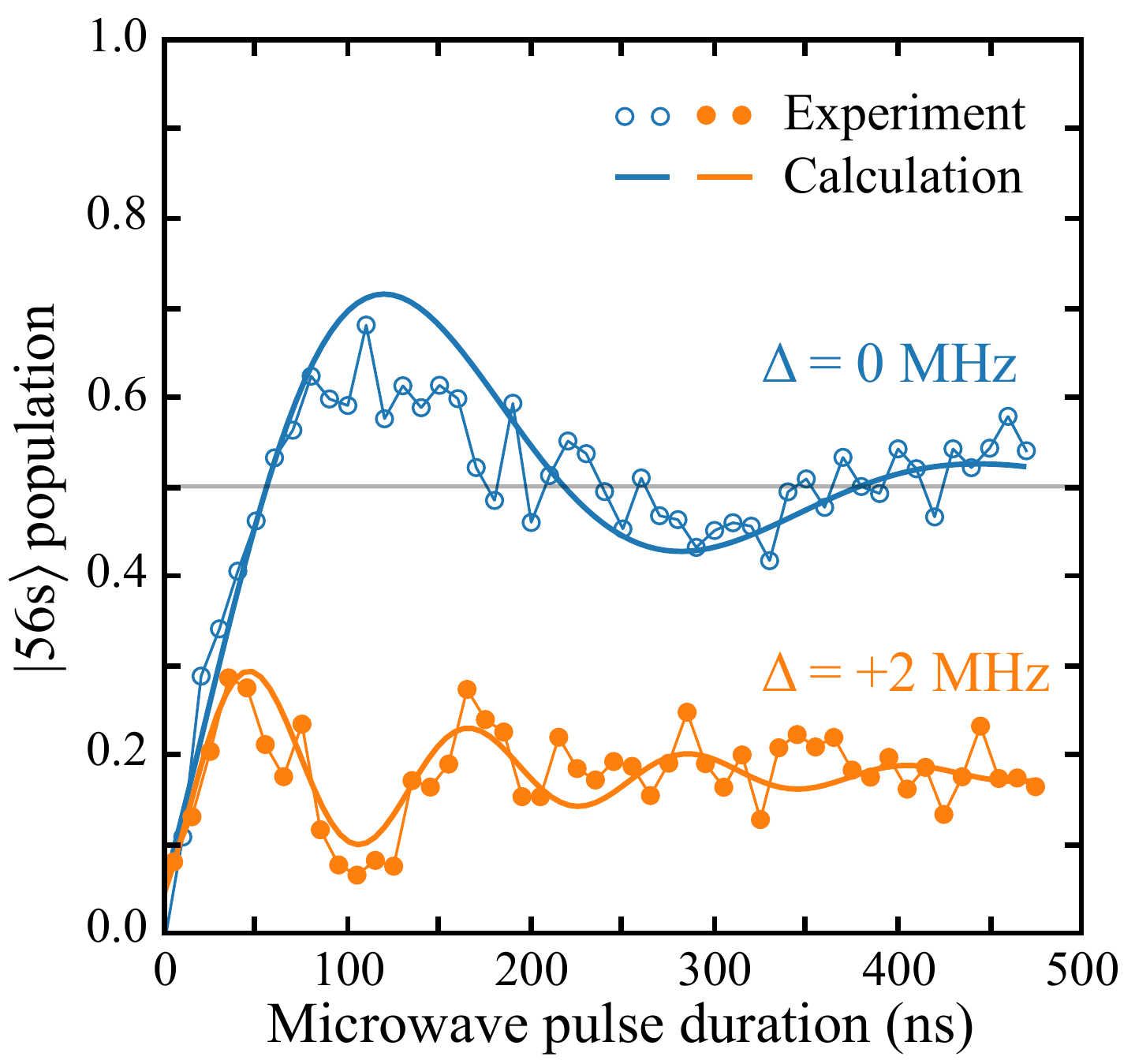}
\caption{Rabi oscillations in the population of the $|56\mathrm{s}\rangle$ state recorded for $P_{\mathrm{source}}=12.6$~mW and the driving microwave field on resonance with, and detuned by $+2$~MHz from, $\nu_{55\mathrm{s},56\mathrm{s}}/2$ as indicated. For both measurements, $T_{\mathrm{CPW}}=3.65$~K.}
\label{fig5}
\end{center}
\end{figure}

The coherence time of the atom--resonator-field interaction was determined by driving Rabi oscillations between the $|55\mathrm{s}\rangle$ and $|56\mathrm{s}\rangle$ states with microwave fields injected into the resonator that were resonant with, $\Delta = 0$~MHz, and detuned by $\Delta=+2$~MHz from $\nu_{55\mathrm{s},56\mathrm{s}}/2$. For these measurements, $T_{\mathrm{CPW}}=3.65$~K, $P_{\mathrm{source}}=12.6$~mW, and the $|56\mathrm{s}\rangle$ signal, normalized to the total Rydberg atom signal, was monitored while the duration of the microwave pulse injected into the resonator was adjusted. For $\Delta=0$~MHz (open circles in Fig.~\ref{fig5}) up to 60\% population transfer to the $|56\mathrm{s}\rangle$ state was observed, but the Rabi oscillation frequency of $\sim3$~MHz permitted the observation of only one complete Rabi cycle within the coherence time of the atom--resonator-field interaction. The coherence time of $\sim150$~ns was inferred by comparison of the experimental data with the overlaid sinusoidal function with an amplitude that decays exponentially with a time constant of 150~ns. To increase the Rabi oscillation frequency a second measurement was made for $\Delta=+2$~MHz. The resulting data (filled circles in Fig.~\ref{fig5}) exhibit an oscillation frequency of $\sim8$~MHz and a similar coherence time of 150~ns (continuous curve). From (1) the $3$~MHz resonant Rabi frequency, (2) the -22.5~dB microwave attenuation between the source and the NbN chip, (3) the microwave field strength in the resonator determined from the data in Fig.~\ref{fig4}(a), and (4) the calculated microwave field distribution above the resonator~\cite{morgan18a}, the typical distance of the atoms from the chip surface in the experiments was estimated to be $\sim100~\mu$m, and the steady-state population of the resonator mode was $\sim10^9$. Within the coherence time of 150~ns, the Rydberg atoms traveled $300~\mu$m along the resonator. Over this distance the microwave field strength and stray electric fields vary. The spatial distribution of the atoms in these inhomogeneous fields represents the dominant source of decoherence. 

In conclusion, we have demonstrated coherent coupling of gas-phase Rydberg atoms to microwave fields in a chip-based superconducting CPW resonator for the first time. By working with He atoms, driving two-photon transitions between the triplet $|55\mathrm{s}\rangle$ and $|56\mathrm{s}\rangle$ Rydberg states -- which exhibit similar static electric dipole polarizabilities -- and using a $\lambda/4$ resonator geometry, effects of charge buildup and stray electric fields~\cite{hattermann12a} were minimized, yielding spectral line widths of the atomic transitions of $\sim1$~MHz. This work can be extended to the single-photon strong-coupling regime by detuning the resonator from the two-photon atomic resonance and dressing the atoms with a strong microwave field with the opposite detuning, or through the use of circular Rydberg states~\cite{morgan18a}. 

\begin{acknowledgments}
We thank Mr John Dumper (UCL) for technical support, Dr Valentina Zhelyazkova (ETH Z\"urich) for contributions to the early stages of the development of the experimental apparatus, Dr Peter Leek (Oxford) and Dr Christoph Zollitsch and Mr Gavin Dold (UCL) for invaluable discussions regarding microwave resonator design and characterization; and Professor Jason Robinson and Mr Chang-Min Lee (Cambridge) for NbN sputtering. This work was supported by the Engineering and Physical Sciences Research Council under Grant No. EP/L019620/1 and through the EPSRC Centre for Doctoral Training in Delivering Quantum Technologies, and the European Research Council (ERC) under the European Union's Horizon 2020 research and innovation program (Grant No. 683341).
\end{acknowledgments}
	

\end{document}